# Reconfiguration of District Heating Network for Operational Flexibility Enhancement in Power System Unit Commitment

Yixun Xue, Mohammad Shahidehpour, *Fellow, IEEE*, Zhaoguang Pan, Bin Wang, Quan Zhou, *Senior Member, IEEE*, Qinglai Guo, *Senior Member, IEEE*, and Hongbin Sun, *Fellow, IEEE*

*Abstract*—Massive adoptions of combined heat and power (CHP) units necessitate the coordinated operation of power system and district heating system (DHS). Exploiting the reconfigurable property of district heating networks (DHNs) provides a cost-effective solution to enhance the flexibility of the power system by redistributing heat loads in DHS. In this paper, a unit commitment considering combined electricity and reconfigurable heating network (UC-CERHN) is proposed to coordinate the day-ahead scheduling of power system and DHS. The DHS is formulated as a nonlinear and mixed-integer model with considering the reconfigurable DHN. Also, an auxiliary energy flow variable is introduced in the formed DHS model to make the commitment problem tractable, where the computational burdens are significantly reduced. Extensive case studies are presented to validate the effectiveness of the approximated model and illustrate the potential benefits of the proposed method with respect to congestion management and wind power accommodation.

## NOMENCLATURE

### A. Sets

| | |
|---|---|
| $T$ | Set of scheduling time index |
| $\kappa^{CHP}/\kappa^{HB}$ | Index set of CHP units/heating boilers |
| $\kappa^{HS}/\kappa^{HES}$ | Set of heat stations/ heat exchange stations |
| $\kappa^{nd}/\kappa^{pipe}$ | Index set of nodes/pipes of the DHS |
| $\kappa^{TU}/\kappa^{WD}$ | Index set of non-CHP units/ wind farms |
| $\kappa^{line}/\kappa^{bus}$ | Index set of transmission lines/buses in power grid |
| $NK_i$ | Index set of extreme points in operational zone of CHP unit $i$ |
| $S_i^{pipe+}/S_i^{pipe-}$ | Index set of pipes flowing from/to node $i$ |
| $S_n^{TU}$ | Index set of non-CHP thermal units with bus $n$ connected |
| $Nd_j^{HS}$ | Index set of node with heat station $i$ connected |
| $Nd_l^{HES}$ | Index set of node with heat exchange station $l$ connected |
| $S_n^{CHP}$ | Index set of CHP units with bus $n$ connected |
| $S_n^{WD}$ | Index set of wind farms with bus $n$ connected |

### B. Parameters and Functions

| | |
|---|---|
| $C_i^{CHP}/C_i^{TU}$ | Cost function of CHP unit/non-CHP thermal unit $i$ |
| $C_i^{HB}$ | Cost function of heating boiler $i$ |
| $C_i^{WD}$ | Cost function of wind curtailment at wind farm $i$ |
| $C_i^{LS}$ | Cost function of load shedding at bus $i$ |
| $D_{n,t}$ | Electrical demand at bus $n$ during period $t$ |
| $F_l$ | Limit of transmission line capacity for line $l$ |
| $\bar{h}_i^{HB}$ | Maximum heating generation limit of heating boiler $i$ |
| $H_{l,t}^{HES}$ | Heat demand of heat exchange station $l$ |
| $L_b/A_b$ | Length/cross sectional area of pipe $b$ |
| $MU_i/RD_i$ | Minimum downtime/uptime of generator $i$ |
| $MR_b$ | Minimum interval of valve operation in pipeline $b$ |
| $\bar{m}_b^{p,s}/\bar{m}_b^{p,r}$ | Maximum mass flow rate limit in pipe $b$ of supply/return networks. |
| $P_i^k/H_i^k$ | Electricity/heat generation with respect to the $k^{th}$ extreme point in the operation region of CHP unit $i$ |
| $\bar{p}_i/\underline{p}_i$ | Minimum/maximum power production of generation unit $i$ |
| $\bar{p}_{i,t}^{WD}$ | Available power generation of wind farm $i$ during period $t$ |
| $RU_i/RD_i$ | Upward/downward power ramping rate of generator $i$ |
| $SU_i/SD_i$ | Startup/shutdown power ramping rate of generator $i$ |
| $SRU_t/SRD_t$ | Upward/downward spinning reserve limit of the whole power system during period $t$ |
| $c/\rho$ | Specific heat capacity /density of water |
| $\lambda_b$ | Heat conductivity coefficient of pipe $b$ |
| $\tau_t^{am}$ | Ground temperature |
| $\underline{\tau}_n^{NS}/\bar{\tau}_n^{NS}$ | Minimum/maximum temperature limit of supply water at node $n$ |

Y. Xue, Z. Pan, B. Wang, Q. Guo, and H. Sun are with the Department of Electrical Engineering, Tsinghua University, China (e-mail: xueyx16@mails.tsinghua.edu.cn, shb@tsinghua.edu.cn). (*corresponding author:Hongbin Sun*)
M. Shahidehpour and Q. Zhou are with Galvin Center for Electricity Innovation, Illinois Institute of Technology, Chicago, IL, USA (e-mail: ms@iit.edu, qzhou15@hawk.iit.edu).

| | |
|---|---|
| $\underline{\tau}_n^{NR}$ / $\overline{\tau}_n^{NR}$ | Minimum/maximum temperature limit of return water at node $n$ |

*C. Variables*

| | |
|---|---|
| $h_{i,t}^{CHP}$ / $h_{i,t}^{HB}$ | Heat output of CHP unit/heating boiler $i$ during period $t$ |
| $p_{i,t}$ | Power production of generator $i$ at period $t$ |
| $p_{i,t}^{WD}$ | Power production of wind farm $i$ at period $t$ |
| $p_{n,t}^{Loss}$ | Load shedding at bus $n$ at period $t$ |
| $\mu_{i,t}$ / $\mu_{b,t}^{R}$ | Binary variable for status of generator $i$ / pipeline $b$, 1 means generation unit/ pipeline are on |
| $x_{i,t}$ / $x_{b,t}^{R}$ | Binary variable, 1 means generation unit $i$ is turned on/ valve at pipeline $b$ is opened |
| $y_{i,t}$ / $y_{b,t}^{R}$ | Binary variable, 1 means generation unit $i$ is turned off/ valve at pipeline $b$ is closed |
| $f_{i,t}^{HB}$ | Fuel consumption of heating boiler $i$ during period $t$ |
| $\alpha_{i,t}^{k}$ | The coefficient of the $k^{th}$ extreme point of CHP thermal unit $i$ during period $t$ |
| $\dot{m}_{b,t}^{PS}$ / $\dot{m}_{b,t}^{PR}$ | Mass flow rate in pipe $b$ of supply/return networks |
| $\dot{m}_{j,t}^{HS}$ / $\dot{m}_{k,t}^{HES}$ | Mass flow rate of heat station $j$ /heat exchange station $k$ during period $t$ |
| $\tau_{n,t}^{NS}$ / $\tau_{n,t}^{NR}$ | Temperature of water flow at node $n$ in supply/return network during period $t$ |
| $\tau_{b,t}^{PS,out}$ / $\tau_{b,t}^{PR,out}$ | Outlet temperature of water flow in pipeline $b$ of supply/return network at period $t$ |
| $\tau_{b,t}^{PS,in}$ / $\tau_{b,t}^{PR,in}$ | Inlet temperature of water flow in pipeline $b$ of supply/return network at period $t$ |

I. INTRODUCTION

By recovery of wasted heat, the combined heat and power (CHP) technology can achieve the cogeneration of electricity and heat with enhanced energy efficiency. CHP units have been broadly utilized around the world. In China, 51% of urban heat demands are supplied by CHP units in winter and the installed capacity of CHP units has reached 550 GW by the end of 2017, which accounts for half of the total installed thermal power capacity [1]. The proliferation of CHP units necessitates the coordinated operation of the power system and district heating system (DHS).

Currently, the power generation of CHP units is mainly determined by heating loads since district heat supply has a more pressing priority than electricity supply, which might pose potential threats to power systems. On one hand, too many CHP units are committed in winter due to a higher heat demand, which would result in reduced energy efficiency and unavoidable wind curtailment, especially in winter midnights with a low electrical load and high heat demand. In Jilin province, Northeastern China, the wind curtailment in winter accounted for 89% of the total curtailment in 2016 [2]. On the other hand, the high proportion of inflexible CHP units would result in limited system ramping capability, which impairs the system operational flexibility and reliability. In fact, the operational flexibility issues (i.e., transmission congestion and balancing challenge) have been reported as the prospective operating challenge of the power system in China, and such issues have been addressed in some researches [3].

Exploiting the flexibility of DHS offers a cost-effective way to cope with the aforementioned issues in power systems. Extensive literature has recently focused on the combined heat and power dispatch (CHPD) problem. Reference [4] characterized the time delays of the heating network based on the node method and exploited the heat storage feature of the DHS. Electric boilers and CHPs were co-optimized to enhance the flexibility of wind integration [5]. The heat transfer process of the extraction steam of the CHP unit was further described using a three-stage model [6]. In [7], four different control modes of DHS were identified and modeled respectively based on the controllability of water temperature and mass flow rates. Also, the flexibility introduced by buildings thermal inertia and heat pumps were explored [8]-[10]. In addition, the hourly unit commitment in combined heat and power system were widely studied. The adjustable robust optimization of CHP units was studied in [11], and a stochastic commitment was modeled in [12]. Reference [13] proposed a short-term unit scheduling with a combined power system and heating networks. In [14], a joint unit commitment of generators and heat-exchanger stations was studied. Reference [15] considered the valve-point effects of CCHP units, where a heuristic algorithm is used to solve the economic environmental commitment. The commitment of the CHP unit system is further investigated in [16] using high-efficient dynamic programming approaches.

The district heating network (DHN) provides a medium to deliver heat from heat sources to different heating loads, which was addressed in the aforementioned studies. However, the configuration of heating networks is usually considered fixed in existing CHPD studies. Actually, similar to the reconfiguration of the distribution power system, the topological structure of DHN is changeable by utilizing remotely controlled valves. The reconfiguration of DHN offers a viable tool in planning and regulation to make sure the quality of heating supply in practical DHS. In Changchun, which is the capital city of Jilin in China, three individual DHSs are connected by tie pipes to realize the DHN reconfiguration operations when needed.

DHN reconfiguration is quite essential for enhancing the reliability of heat supply. Reference [17] simulated the reconfiguration method for the ring-shaped heating supply network in terms of fault conditions. In [18], a planning model of pipe connections among different heating supply systems was developed based on the limited-heating requirement. Besides, the reconfigurable feature of DHS is also utilized to enhance the efficiency of heating supply. In [19], consideration was given to the problem of changing the configuration for heat supply with reservation of pipelines and a definition of the limiting graph for pipeline reconfiguration was proposed. The methodology is presented in [20] for the optimization-based reconfiguration of heat-exchanger networks.

However, the DHN reconfiguration has not been considered in the coordinated operation of power system and DHS for enhancing the overall system performance. On one hand, the DHN reconfiguration can redistribute heat loads among participating heat sources for reducing heat losses and heat supply costs in DHS. On the other hand, the DHN reconfiguration can enhance the operational flexibility of CHP units through valve regulations, which has great potential for wind integration and congestion alleviation in power system. Accordingly, this paper aims to develop a day-ahead unit scheduling methodology with considering the crucial reconfigurable features of heating networks that provide commitments of generation units and regulations of valves. The contributions of this paper are summarized as follows:

1) The reconfigurable DHS models and the operation of valves in day ahead-scheduling are integrated into the unit commitment problem of power system, where the system energy efficiency and operational flexibility can be enhanced via proper DHN reconfigurations. To the best of our knowledge, this is the first paper to address the DHN reconfiguration when conducting a CHPD problem.

2) To characterize the reconfigurable property of DHN, the exact DHS model considering valve operation is proposed in this paper, which is formulated as a mixed-integer nonlinear problem. Also, a generalized energy flow model is developed in the paper by approximating heat loss for reducing the computational burdens.

3) The potential benefits of DHN reconfiguration in terms of congestion elimination and wind power integration are validated and illustrated using both simulated and actual systems.

The remainder is organized as follows: The exact model for DHN reconfiguration and corresponding energy flow model is developed in Section II. In Section III, based on the developed linear energy flow model, the unit commitment considering combined electricity and reconfigurable heating networks (UC-CERHN) is formulated. In Section IV, case studies are presented to validate the effectiveness of the proposed method. Section V concludes.

## II. MODELING THE RECONFIGURABLE DHS

### A. DHS Structure

The DHS is formed from the heat stations, heat exchange stations, pipeline networks, and heat loads. Similar to the structure of the transmission and distribution systems in the power system, the pipelines in DHS can be classified into primary networks and secondary networks, as shown in Fig1.

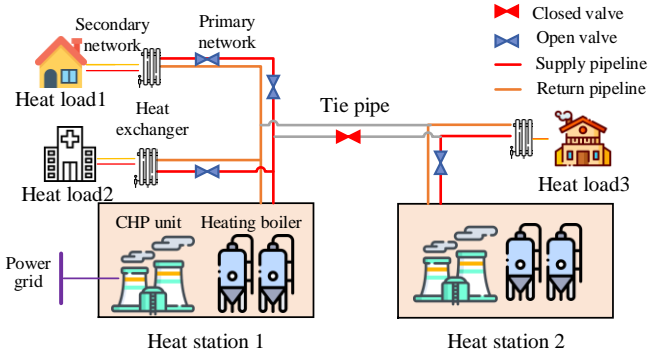

Fig. 1. The configuration of the DHS.

Heat stations equipped with heat sources (e.g., CHP units and heating boilers) supply heat for the DHN. Heat is generated by heat sources, delivered by circulating hot water flow through the primary network, and procured by the heat exchangers. The secondary networks deliver heat from heat exchange stations to every heat load. In Fig. 1, the primary networks and secondary networks are connected via heat exchangers. The configuration of the primary network can be changed by the remote control of valves on the available sectionalizing and tie pipes [17]. The topology of secondary networks is considered fixed in field practice since there are few measuring points and adjustable valves in secondary networks.

In this paper, we focus on the reconfiguration of the primary networks, in which the exact model for reconfigurable DHS is presented and approximated to an energy flow model for reducing the computational burden.

### B. Exact Model for DHS

#### 1) Heat Station

In heat stations, CHP units and heating boilers amounted for 93% of the total heat sources in China are considered [1]. Using the convex combination of extreme points, the electricity and heat generation of CHP units are modeled as [4]:

$$p_{i,t} = \sum_{k=1}^{NK_i} \alpha_{i,t}^k P_i^k, \forall i \in \kappa^{CHP}, t \in T, \quad (1)$$

$$h_{i,t}^{CHP} = \sum_{k=1}^{NK_i} \alpha_{i,t}^k H_i^k, \forall i \in \kappa^{CHP}, t \in T, \quad (2)$$

$$\sum_{k=1}^{NK_i} \alpha_{i,t}^k = 1, 0 \leq \alpha_{i,t}^k \leq 1, \quad (3)$$
$$\forall i \in \kappa^{CHP}, k \in \{1,2,...,NK_i\}, t \in T.$$

The heating boiler is usually utilized to offer auxiliary heat supplement due to its high generation costs and quick regulating capacity, which is modeled as:

$$h_{i,t}^{HB} = \eta_i^{HB} \cdot f_{i,t}^{HB}, \forall i \in \kappa^{HB}, t \in T. \quad (4)$$

$$0 \leq h_{i,t}^{HB} \leq \overline{h}_i^{HB}, \forall i \in \kappa^{HB}, t \in T. \quad (5)$$

The heat generation of the heat station, including the heat production of both heating boilers and CHP units, is utilized for warming up the mass flow from the return pipes of the DHS, stated as:

$$\sum_{i \in \kappa_j^{CHP}} \mu_{g,t} h_{i,t}^{CHP} + \sum_{i \in \kappa_j^{HB}} h_{i,t}^{HB} = c \cdot \dot{m}_{j,t}^{HS} \cdot (\tau_{n,t}^{NS} - \tau_{n,t}^{NR}), \quad (6)$$
$$\forall j \in \kappa^{HS}, n = Nd_j^{HS}, t \in T.$$

$$\underline{\tau}_n^{NS} \leq \tau_{n,t}^{NS} \leq \overline{\tau}_n^{NS}, \forall n = Nd_i^{HS}, t \in T. \quad (7)$$

#### 2) Heat Exchange Stations

Heat exchange station is regarded as a heat load of the primary network, stated as:

$$c \cdot \dot{m}_{l,t}^{HES} \cdot (\tau_{n,t}^{NS} - \tau_{n,t}^{NR}) = H_{l,t}^{HES}, \forall l \in \kappa^{HES}, n = Nd_l^{HES}, t \in T. \quad (8)$$

$$\underline{\tau}_n^{NR} \leq \tau_{n,t}^{NR} \leq \overline{\tau}_n^{NR}, \forall l \in \kappa^{HES}, n = Nd_l^{HES}, t \in T \quad (9)$$

#### 3) Reconfigurable Primary Heating Network

Different from the existing model in [13], both hydraulic regimes and thermal conditions of DHN are considered variable when the primary network is reconfigurable. Thus, the continuity of mass flow, temperature mixing, and the heat loss of water should be considered. Also, the direction of water flow in primary pipes can be changed when the configuration of the primary network changes.

The total mass flow rate of water entering every node is zero based on the continuity of mass flow:

$$\sum_{b\in S_i^{pipe-}} \mu_{b,t} \dot{m}_{b,t}^{PS} + \sum_{j\in \kappa_i^{HS}} \dot{m}_{j,t}^{HS} = \sum_{k\in \kappa_i^{HES}} \dot{m}_{k,t}^{HES} + \sum_{b\in S_i^{pipe+}} \mu_{b,t} \dot{m}_{b,t}^{PS}, \quad (10)$$

$$\sum_{b\in S_i^{pipe+}} \mu_{b,t} \dot{m}_{b,t}^{PR} + \sum_{j\in \kappa_i^{HS}} \dot{m}_{j,t}^{HS} = \sum_{k\in \kappa_i^{HES}} \dot{m}_{k,t}^{HES} + \sum_{b\in S_i^{pipe-}} \mu_{b,t} \dot{m}_{b,t}^{PR}. \quad (11)$$

$$-\overline{\dot{m}}_b^{p,s} \le \dot{m}_{b,t}^{p,s} \le \overline{\dot{m}}_b^{p,s}, \forall b \in \kappa^{pipe}, t\in T, \quad (12)$$

$$-\overline{\dot{m}}_b^{p,r} \le \dot{m}_{b,t}^{p,r} \le \overline{\dot{m}}_b^{p,r}, \forall b \in \kappa^{pipe}, t\in T, \quad (13)$$

where $\dot{m}_{b,t} < 0$ indicates that the actual direction of water flow in pipeline $b$ is opposite to the reference direction.

Mass flow with different temperature is mixed at confluence nodes. The temperature mixing of water is stated as:

$$\sum_{b\in S_i^{pipe-}} (\tau_{b,t}^{PS,out} \mu_{b,t} \dot{m}_{b,t}^{PS}) = \tau_{i,t}^{NS} \cdot \sum_{b\in S_i^{pipe+}} \mu_{b,t} \dot{m}_{b,t}^{PS}, \forall i\in \kappa^{nd}, t\in T, \quad (14)$$

$$\sum_{b\in S_i^{pipe-}} (\tau_{b,t}^{PR,out} \mu_{b,t} \dot{m}_{b,t}^{PR}) = \tau_{i,t}^{NR} \cdot \sum_{b\in S_i^{pipe+}} \mu_{b,t} \dot{m}_{b,t}^{PR}, \forall i\in \kappa^{nd}, t\in T, \quad (15)$$

$$\tau_{b,t}^{PS,in} = \tau_{i,t}^{NS}, \forall i\in \kappa^{nd}, b\in S_i^{pipe+}, t\in T, \quad (16)$$

$$\tau_{b,t}^{PR,in} = \tau_{i,t}^{NR}, \forall i\in \kappa^{nd}, b\in S_i^{pipe-}, t\in T. \quad (17)$$

Noting that the temperature mixing equations (14)-(17) hold only if the inlet and outlet of pipelines are determined correctly (i.e., actual directions of pipelines connected to node $i$ are consistent with the reference directions). Otherwise, water flow may not be mixed at node $i$, and equations (14)-(17) should be reformulated. However, in a reconfigurable DHN, the mass flow directions cannot be predetermined, which brings a great challenge for adopting the exact DHS model in the UC-CERHN problem. Besides, the temperature declines gradually from the pipeline to the ambient environment. The heat loss is accounted as:

$$\tau_{b,t}^{PS,out} = \tau_t^{am} + (\tau_{b,t}^{PS,in} - \tau_t^{am}) e^{-\frac{\lambda_b L_b}{\dot{m}_{b,t}^{PS} A_b \rho c}}, \forall b \in \kappa^{pipe}, t\in T, \quad (18)$$

$$\tau_{b,t}^{PR,out} = \tau_t^{am} + (\tau_{b,t}^{PR,in} - \tau_t^{am}) e^{-\frac{\lambda_b L_b}{\dot{m}_{b,t}^{PR} A_b \rho c}}, \forall b \in \kappa^{pipe}, t\in T. \quad (19)$$

It can be demonstrated that equations (18)-(19) still hold when $\dot{m}_{b,t}^{PR}$ and $\dot{m}_{b,t}^{PS}$ are negative.

In addition, the configuration of the primary network cannot be changed frequently, which will lead to the instability of the pipeline:

$$\mu_{b,t} - \mu_{b,t-1} = x_{b,t}^R - y_{b,t}^R, \forall b \in \kappa^{pipe}, t\in T, \quad (20)$$

$$\sum_{\sigma=\max\{1,t-MR_b+1\}}^{t} x_{b,\sigma}^R \le \mu_{b,t}^R, \forall b \in \kappa^{pipe}, t\in T, \quad (21)$$

$$\sum_{\sigma=\max\{1,t-MR_b+1\}}^{t} y_{b,\sigma}^R \le 1-\mu_{b,t}^R, \forall b \in \kappa^{pipe}, t\in T. \quad (22)$$

The multi-source operation of DHS could lead to difficulty in pressure balance and the make-up of the water. Thus, different heat stations in DHS are typically operated in isolation in field practice when heating network is reconfigured [18], which is modeled as:

$$\sum_{b\in \kappa^{pipe}} \mu_{b,t}^R = \sum_{b\in \kappa^{pipe}} \tilde{\mu}_{b,t}^R, \forall t\in T. \quad (23)$$

which ensures that the sum of statuses of all pipelines will not be changed after reconfiguration.

### C. An Approximate Heat Flow Model

The exact DHS model is nonlinear and there are many integer variables owing to the incorporation of DHN reconfiguration, which might lead to intractability for solving the UC-CERHN problem. Moreover, the temperature mixing equations (14)-(17) cannot be predetermined due to the unknown mass flow direction. Thus, the original DHS model cannot be applied directly, which should be transformed as an energy flow model.

#### 1) Reformation

Hot water in pipelines is a medium for heat delivery. Thus, the hot water flow in pipelines of the DHSs actually represents the transference of thermal energy. The available heat quantity in the pipeline denoted by $h_b$ is defined as:

$$h_b = c\dot{m}_b (\tau_b^{PS} - \tau_b^{PR}), \forall b \in \kappa^{pipe}. \quad (24)$$

Using (24), the bilinear terms (i.e., the product of temperature and mass flow rate) in (6), (14), and (15) are replaced by auxiliary heat quantity variables and theses nonlinear equations can be reformulated as linear constraints (25)-(26). Also, the heat loss of water can be rewritten as (27)-(28). The heat quantity is bounded based on equations (7), (9), and (12)-(13), which is stated in (29)-(30).

$$\sum_{i\in \kappa_j^{CHP}} h_{i,t}^{CHP} + \sum_{i\in \kappa_j^{HB}} h_{i,t}^{HB} = h_{j,t}^{HS}, \forall j\in \kappa^{HS}, t\in T, \quad (25)$$

$$\sum_{b\in S_i^{pipe-}} h_{b,t}^{P,out} + \sum_{j\in S_i^{HS}} h_{j,t}^{HS} = \sum_{k\in S_i^{HES}} h_{k,t}^{HES} + \sum_{b\in S_i^{pipe+}} h_{b,t}^{P,in}, \quad (26)$$

$$h_{b,t}^{P,out} = h_{b,t}^{P,in} - h_{b,t}^{P,loss}, \forall b \in \kappa^{pipe}, t\in T, \quad (27)$$

$$h_{b,t}^{P,loss} = c\dot{m}_{b,t}^{PS} (\tau_{b,t}^{PS,in} + \tau_{b,t}^{PR,in} - 2\tau_t^{am})(1 - e^{-\frac{\lambda_b L_b}{\dot{m}_{b,t}^{PS} A_b \rho c}}), \quad (28)$$

$$\forall b \in \kappa^{pipe}, t\in T,$$

$$\mu_{i,t} \underline{h}_i^{CHP} \le h_{i,t}^{CHP} \le \mu_{i,t} \overline{h}_i^{CHP}, \forall i\in \kappa^{CHP}, t\in T \quad (29)$$

$$-\mu_{b,t} \overline{h}_b^P \le h_{b,t}^{P,out} \le \mu_{b,t} \overline{h}_b^P, -\mu_{b,t} \overline{h}_b^P \le h_{b,t}^{P,in} \le \mu_{b,t} \overline{h}_b^P. \quad (30)$$

The above energy flow model (25)-(30) is consistent with the energy conservation law in DHS, which is related to heat quantity variables (e.g., $h_{b,t}^{P,out}$, $h_{b,t}^{P,in}$), temperature and mass flow rate variables.

#### 2) Approximation of the Heat Loss

If the heat loss is determined, the energy flow model in (25)-(30) is independent of temperature and mass flow rate variables, which can be solved directly without considering the direction of water flow. According to (28), the heat loss in every pipeline is characterized by an exponential function about the water temperature and mass flow rates. The nonlinear term $\lambda_b L_b / \dot{m}_{b,t}^{PS} A_b \rho c$ in (28) is close zero, which is smaller than 0.001 in general. Considering $1 - e^{-x} \approx x$, the heat loss is approximated by:

$$h_{b,t}^{P,loss} \approx c\dot{m}_{b,t}^{PS} (\tau_{b,t}^{PS,in} + \tau_{b,t}^{PR,in} - 2\tau_t^{am}) \frac{\lambda_b L_b}{\dot{m}_{b,t}^{PS} A_b \rho c}$$
$$= \frac{\lambda_b L_b}{A_b \rho} (\tau_{b,t}^{PS,in} + \tau_{b,t}^{PR,in} - 2\tau_t^{am}), \forall b \in \kappa^{pipe}, t\in T. \quad (31)$$

Equation (31) is the first-order Taylor expansion of (28), which suggests that $h_{b,t}^{P,loss}$ is independent of mass flow rate. The phenomenon has been validated by a series of experiments in practical DHS recently [21], which is because the heat loss in pipes is mainly caused by heat dissipation to the environment.

The water temperature of supply and return networks are bounded. For example, the temperature of mass flow in supply pipes is usually bounded between 80-100°C. As the sole controllable variable that affects heat loss in pipes is the inlet temperature, the DHN should be controlled to keep the operation temperature as low as possible to cut down the heat loss. Here, the inlet temperature in (31) is set as its lower limits when calculating the heat loss, and $h_{b,t}^{P,loss}$ can be stated as:

$$h_{b,t}^{P,loss} = \frac{\lambda_b L_b}{A_b \rho}(\underline{\tau}_b^{PS,in} + \underline{\tau}_b^{PR,in} - 2\tau_t^{am}), \quad \forall b \in \kappa^{pipe}, t \in T \quad (32)$$

Compared to practical heat loads in DHS, the heat loss along pipelines is relatively small [7]. Then, the errors corresponding to the constant heat loss approximation would be negligible, which will be illustrated in section IV. Accordingly, the energy flow model is transformed into a mixed-integer linear model, which is independent of temperature and mass flow rate variables. It implies that the UC-CHEN problem can be solved with significantly reduced computational burdens while without loss of accuracy.

Once the energy flow of the DHN $h_b^{sp}$ is determined, the detailed temperature and mass flow rate variables can be calculated using the hydraulic and thermal equations of the DHS, which are formulated as the following compact form:

$$\begin{cases} Ax_m = 0, Bx_\tau = 0, \\ g(x_m, x_\tau) = 0, h_b^{sp} = h(x_m, x_\tau). \end{cases} \quad (33)$$

where $Ax_m = 0$ refers to equations (10)-(11), $Bx_\tau = 0$ refers to equations (16)-(17), $g(x_m, x_\tau) = 0$ refers to equations (14)-(15), (18)-(19), and $h_b^{sp} = h(x_m, x_\tau)$ refers to equation (24). Variables of the mass flow rate are denoted as $x_m$, water temperature variables are denoted as $x_\tau$, and $h_b^{sp}$ represents the heat quantity variables, including $h_{b,t}^{P,out}$, $h_{b,t}^{P,in}$, $h_{j,t}^{HS}$, and $h_{k,t}^{HES}$. The proposed linear energy flow model is much tractable when variable flow, variable temperature, and DHN reconfiguration are considered.

## III. UC-CERHN FORMULATION

The UC-CERHN problem is formulated in this section to coordinate the day-ahead scheduling of power systems and DHSs, including the commitment of CHP units and non-CHP thermal units, and switching operation of valves. The energy flow model (33) is applied to solve the UC-CERHN.

### A. Objective

The objective of the proposed UC-CREHN problem is the total operation cost of the power system and district heating system:

$$obj = \sum_{t \in T} \left( \sum_{i \in \kappa^{CHP}} C_i^{CHP}(p_{i,t}, h_{i,t}^{CHP}, u_{i,t}, x_{i,t}, y_{i,t}) + \sum_{i \in \kappa^{HB}} C_i^{HB}(h_{i,t}^{HB}) \right. \\ \left. + \sum_{i \in \kappa^{TU}} C_i^{TU}(p_{i,t}, u_{i,t}, x_{i,t}, y_{i,t}) + \sum_{i \in \kappa^{WD}} C_i^{WD}(p_{i,t}^{WD}) + \sum_{i \in \kappa^{bus}} C_i^{LS}(p_{i,t}^{Loss}) \right) \quad (34)$$

The objective in (34) includes the operation cost from CHP units, heating boilers, and the non-CHP thermal units, and the penalty cost from wind power curtailment and power load shedding. The operation cost functions of CHP units, heating boilers, and non-CHP thermal units are convex quadratic functions[10].

### B. Constraints

The UC-CREHN is subject to the operation constraints of the power system and reconfigurable district heating system. The DHS constraints are presented in (1)-(5), energy flow model is presented in (25)-(27),(29)-(30), and the commitment constraints of pipelines are presented in (20)-(23). The operation constraints of the power system are stated as:

$$\sum_{i \in \kappa^{TU} \cup \kappa^{CHP}} p_{i,t} + \sum_{i \in \kappa^{WD}} p_{i,t}^{WD} = \sum_{n \in \kappa^{bus}} (D_{n,t} - p_{n,t}^{Loss}), t \in T, \quad (35)$$

$$\left| \sum_{n \in \kappa^{bus}} SF_{l,n} \cdot \left( \sum_{i \in S_n^{TU} \cup S_n^{CHP}} p_{i,t} + \sum_{i \in S_n^{WD}} p_{i,t}^{WD} - D_{n,t} + p_{n,t}^{Loss} \right) \right| \quad (36)$$
$$\leq F_l, \forall l \in \kappa^{line}, t \in T,$$

$$u_{i,t} \underline{p}_i \leq p_{i,t} \leq u_{i,t} \bar{p}_i, \forall i \in \kappa^{TU} \cup \kappa^{CHP}, t \in T, \quad (37)$$

$$0 \leq p_{i,t}^{WD} \leq \bar{p}_i^{WD}, \forall i \in \kappa^{WD}, t \in T, \quad (38)$$

$$-(1-u_{i,t+1})SD_i - RD_i u_{i,t+1} \leq p_{i,t+1} - p_{i,t},$$
$$p_{i,t+1} - p_{i,t} \leq (1-u_{i,t})SU_i + RU_i u_{i,t}, \quad (39)$$
$$\forall i \in \kappa^{TU} \cup \kappa^{CHP}, t \in T,$$

$$\sum_{i \in \kappa^{TU}} \min\{RU_i, u_{i,t}\bar{p}_i - \bar{p}_{i,t}\} \geq SRU_t, \forall i \in \kappa^{TU}, t \in T, \quad (40)$$

$$\sum_{i \in \kappa^{TU}} \min\{RD_i, \bar{p}_{i,t} - u_{i,t}\underline{p}_i\} \geq SRD_t, \forall i \in \kappa^{TU}, t \in T, \quad (41)$$

$$u_{i,t} - u_{i,t-1} = x_{i,t} - y_{i,t-1}, \forall i \in \kappa^{TU} \cup \kappa^{CHP}, \forall t \in T, \quad (42)$$

$$\sum_{k=\max\{1,t-MU_i+1\}}^{t} x_{i,k} \leq u_{i,t}, \forall i \in \kappa^{TU} \cup \kappa^{CHP}, t \in T, \quad (43)$$

$$\sum_{k=\max\{1,t-MD_i+1\}}^{t} y_{i,k} \leq 1 - u_{i,t}, \forall i \in \kappa^{TU} \cup \kappa^{CHP}, t \in T. \quad (44)$$

The power system operation subjects to the power balance constraint (35), transmission line capacity constraints (36), unit generation limits (37)-(38), ramping limit (39), spinning reserve limits (40)-(41), logic constraint of generation unit status (42), and the constraints of minimum down/uptime of generation units in (43)-(44). The UC-CERHN is a mixed-integer linear model with quadratic objective, which can be effectively solved using our previous work [14].

### C. Discussion of Flexibility from DHN Reconfiguration

DHN delivers heat from heat stations to different loads, which determines the allocation of heat sources. In UC-CERHN model, the heat station supplies the heat demand of its connected heating loads and heat losses, stated as:

$$h_{j,t}^{HS} = \sum_{k \in \kappa_j^{HES}} h_{k,t}^{HES} + \sum_{b \in \kappa^{pipe}} h_{b,t}^{P,loss}, \forall\ j \in \kappa^{HS}, t \in T. \quad (45)$$

From the perspective of DHS, when DHN reconfiguration occurs, the heat load would be reallocated among different heat stations and heat loss varies (i.e., the right side of equation (45) changes). Therefore, the operation cost of DHS (the first two items in equation (34)) can be reduced due to the utilization of cheaper heat stations and lower heat losses.

Meanwhile, the commitment statuses and outputs of heat sources (the left side of equation (45)) in different heat stations can be effectively regulated via DHN reconfiguration, implying that the power output of CHP units can be regulated with enhanced flexibility. Therefore, the DHN reconfiguration provides a promising solution to promote wind power integration and congestion alleviation in power system, which looks particularly appealing since it allows power system operators to reduce the operation cost (the latter two items in equation (34)) by valve operations without additional costly generation or load shedding.

## IV. CASE STUDIES

### A. Test System Configuration

Fig. 2 depicts a test system composed of a 6-bus EPS and an 8-node DHS, which is named the P6H6 system. The heat load in the DHS is supplied by two heat stations (HS1 and HS2). HS1 is equipped with one heating boiler and one extraction-condensing CHP unit (CHP1), which is connected to B3 in the power system. The back-pressure CHP unit (CHP2) is the only heat source in HS2. In the business as usual network operation, the sectionalizing valves (v1-v4, v6-v7) are normally open and the tie valve (v5) is normally closed, i.e., there is no mass flow for pipe N3-N7 normally.

The system-wide upward/downward spinning reserve requirements of the power system are set as 40 MW. Fig. 3 shows the profiles of total load and the wind power forecast [22]. A day-ahead hourly UC-CERHN is solved using the Gurobi Solver interfaced through MATLAB.

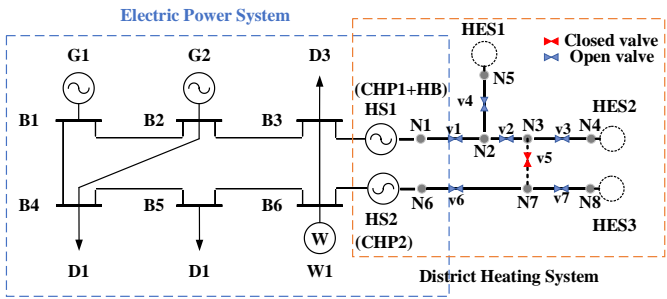

Fig. 2. Configuration of the P6H6 system.

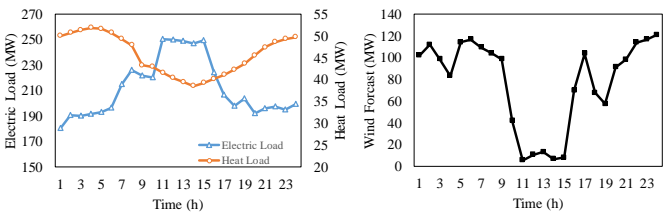

Fig. 3. Total load and wind forecast profile of P6H6 system.

### B. Commitment Result & Analysis

To investigate the effect of heating network reconfiguration on commitment result, the following two cases are conducted:

*Case 1* (base case): the DHN reconfiguration is ignored. The DHS operates in the usual configuration and v5 remains closed. In this case, heat load in HES3 is always supplied by HS2, and HES1-HES2 are supplied by HS1.

*Case 2*: the heating networks is reconfigurable by the remote control of the available sectionalizing and tie valves.

TABLE I
COMPARISON OF OPERATION COSTS IN CASE 1 AND CASE 2

| | $\sum_{t \in T}\sum_{i \in \kappa^{TU}} C_{i,t}^{TU}$ | $\sum_{t \in T}\sum_{i \in \kappa^{CHP}} C_{i,t}^{CHP}$ | $\sum_{t \in T}\sum_{i \in \kappa^{WD}} C_{i,t}^{WD}$ | $\sum_{t \in T}\sum_{i \in \kappa^{HB}} C_{i,t}^{HB}$ | Total cost |
|---|---|---|---|---|---|
| Case1 | 48659 | 38248 | 4100 | 7496 | 98503 |
| Case2 | 49068 | 32079 | 2227 | 10037 | 93411 |

TABLE II
VALVE OPERATION FOR HEATING NETWORK RECONFIGURATION

| Critical period | 1 | 3 | 9 | 13 | 16 | 20 | 23 |
|---|---|---|---|---|---|---|---|
| Open valves | v5 | v6 | v5 | v2 | v1 | v2 | v5 |
| Closed valves | v6 | v5 | v2 | v1 | v2 | v5 | v6 |
| Mass flow | N3→N7 | N6→N7 | N7-N3 | N3→N2 | - | N2→N3 | N3→N7 |

Table I compares the operation costs in Case 1 and Case 2. Table II shows the valve operation scheduling for heating network reconfigurable over 24 hours. The result analyses are stated as follows.

1) In Case 2, the total production cost decreases by 5.17% and the heating cost (i.e., the sum of $\sum_{t \in T}\sum_{i \in \kappa^{CHP}} C_{i,t}^{CHP}$ and $\sum_{t \in T}\sum_{i \in \kappa^{HB}} C_{i,t}^{HB}$) is reduced by 7.93%, as compared with that of Case 1. It implies that heat demand is supplied in a more economical manner due to the effective DHN reconfiguration.

Figs. 4 and 5 show the generation schedule in Case 1 and Case 2, respectively. The heat generation of CHP2 increases greatly in periods 9-19 in Case 2 compared to that in case1. That is because the switching actions of valves v2 and v5, and the HES2 is transferred to be supplied by the CHP2, a less-expensive CHP unit, in periods 9-12 and 16-19. In periods 13-15, when the total heat load is lowest, HES1 will also be redistributed to CHP2. In Table IV, CHP1 is not committed in periods 9-19 due to its high generation cost, and the DHN reconfiguration enables CHP2 to supply more heat loads.

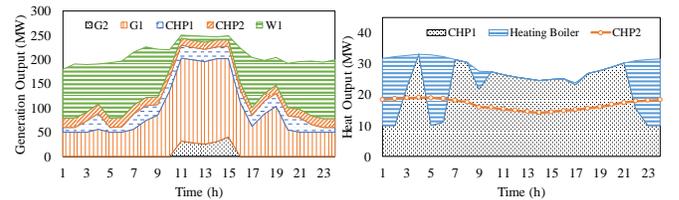

Fig. 4. Unit schedule in Case 1: (a) Power output of all units. (b) Heat generation.

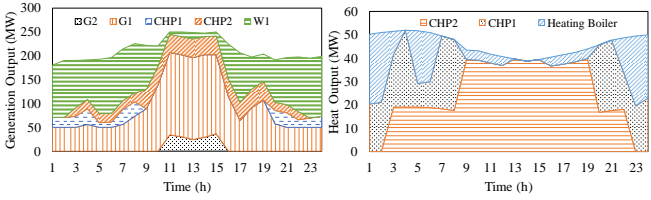

Fig. 5. Unit schedule in Case 2: (a) Power output of all units. (b) Heat generation.

2) The heat load can be shifted by the DHN reconfiguration so that the power production of the CHP units can be regulated for better wind accommodation. In Table II, v5 is switched on at periods 1-2 and 23-24 in Case 2 when wind generation is relatively high. The heat load in HES3 is redistributed to be supplied by HS1 while CHP2 is not committed during these night periods. Thus, the heating boiler in HS1 can be fully utilized to produce more heat and reduce the output of CHP units, which can spare more space for wind power integration during these periods. In Table I, the wind curtailment decreased by 51.8% in Case 2 as compared to that in Case 1.

3) The commitment status of generations is interdependent with the switching actions of valves. Tables III and IV give the unit commitment results in Case 1 and Case 2. In Case 1, both CHP1 and CHP2 remain on at all hours to guarantee the heat supply of corresponding heat loads. In Case 2, due to the increased flexibility introduced by DHN reconfiguration, the CHP1 is not committed at off-peak hours of heat load and CHP2 is terminated during some night period, which can effectively alleviate the wind curtailment and reduce the total operation cost.

TABLE III
UNIT COMMITMENT IN CASE 1

| Hour | 1 | 2 | 3 | 4 | 5 | 6 | 7 | 8 | 9 | 10 | 11 | 12 | 13 | 14 | 15 | 16 | 17 | 18 | 19 | 20 | 21 | 22 | 23 | 24 |
|---|---|---|---|---|---|---|---|---|---|---|---|---|---|---|---|---|---|---|---|---|---|---|---|---|
| G1 | 1 | 1 | 1 | 1 | 1 | 1 | 1 | 1 | 1 | 1 | 1 | 1 | 1 | 1 | 1 | 1 | 1 | 1 | 1 | 1 | 1 | 1 | 1 | 1 |
| G2 | 0 | 0 | 0 | 0 | 0 | 0 | 0 | 0 | 0 | 0 | 1 | 1 | 1 | 1 | 1 | 1 | 0 | 0 | 0 | 0 | 0 | 0 | 0 | 0 |
| CHP1 | 1 | 1 | 1 | 1 | 1 | 1 | 1 | 1 | 1 | 1 | 1 | 1 | 1 | 1 | 1 | 1 | 1 | 1 | 1 | 1 | 1 | 1 | 1 | 1 |
| CHP2 | 1 | 1 | 1 | 1 | 1 | 1 | 1 | 1 | 1 | 1 | 1 | 1 | 1 | 1 | 1 | 1 | 1 | 1 | 1 | 1 | 1 | 1 | 1 | 1 |

TABLE IV
UNIT COMMITMENT IN CASE 2

| Hour | 1 | 2 | 3 | 4 | 5 | 6 | 7 | 8 | 9 | 10 | 11 | 12 | 13 | 14 | 15 | 16 | 17 | 18 | 19 | 20 | 21 | 22 | 23 | 24 |
|---|---|---|---|---|---|---|---|---|---|---|---|---|---|---|---|---|---|---|---|---|---|---|---|---|
| G1 | 1 | 1 | 1 | 1 | 1 | 1 | 1 | 1 | 1 | 1 | 1 | 1 | 1 | 1 | 1 | 1 | 1 | 1 | 1 | 1 | 1 | 1 | 1 | 1 |
| G2 | 0 | 0 | 0 | 0 | 0 | 0 | 0 | 0 | 0 | 0 | 1 | 1 | 1 | 1 | 1 | 1 | 0 | 0 | 0 | 0 | 0 | 0 | 0 | 0 |
| CHP1 | 1 | 1 | 1 | 1 | 1 | 1 | 1 | 0 | 0 | 0 | 0 | 0 | 0 | 0 | 0 | 0 | 0 | 0 | 1 | 1 | 1 | 1 | 1 | 1 |
| CHP2 | 0 | 0 | 1 | 1 | 1 | 1 | 1 | 1 | 1 | 1 | 1 | 1 | 1 | 1 | 1 | 1 | 1 | 1 | 1 | 1 | 1 | 1 | 0 | 0 |

### C. Effect of Heating Network Reconfiguration on Congestion Management

To show that DHN reconfiguration facilitates eliminating congestion in the electricity network, Case 1 and Case 2 are compared under different transmission capacity of line B2-B4, which is a heavily loaded line in the P6H6 system. The results are summarized in Table V, where the load shedding, power generation of CHP units at period 15 (peak time of electric load), and total cost of UC-CERHN for two cases are given. The shift factor for CHP 1 and CHP2 on line B2-B4 are -0.3422 and 0.1242, respectively.

Here, three scenarios are presented to illustrate the effect of DHN reconfiguration on congestion management, where the transmission capacity of line B2-B4 is set as 100MW, 90MW, and 80MW in Scenarios 1, 2, and 3 respectively. In Scenario 1, the corresponding power flow from B2-B4 in Case 1 is 89.21 MW and that in Case 2 is 92.34 MW, both of which are below the capacity and total cost of Case 2 is below that in Case 2. In Scenario 2, the DHN in Case 2 have same configuration for heat supply as Case 1 in peak load periods with v5 closed and v6 opened. The heat load in HES1 and HES2, which are originally supplied by HS2 in baseline of Case 2, are shifted to be supplied by HS1. Thus, the CHP1 generates more and power output of CHP2 reduces, and the congestion in line B1-B4 is prevented thorough the reallocation of CHP generation. In Scenario 3, the load shedding of power system is inevitable in Case 1 for alleviating transmission congestion of line B2-B4. Comparatively, in Case 2, all heat load will be supplied by HS1 thorough DHN reconfiguration and CHP1 is committed to generate at its maximum power output, where the transmission congestion is also alleviated without load shedding. In Table V, as the transmission capacity of line B2-B4 decreases from Scenario 1 to 3, the overload in line B2-B4 can be alleviated via proper DHN reconfiguration while load shedding is avoided.

TABLE V
COMPARISON OF CASE 1 AND CASE 2 IN DIFFERENT CAPACITY OF LINE B2-B4

| Case | items | Transmission capacity of line B2-B4 (MW) | | |
|---|---|---|---|---|
| | | 100 | 90 | 80 |
| Case1 | Load shedding (MW) | 0 | 0 | 14.95 |
| | CHP generation (MW) (CHP1,CHP2) | 39.4 (25,14.4) | 39.4 (25,14.4) | 42.9 (28.5,14.4) |
| | Total cost ($) | 98503 | 98503 | 100547 |
| Case2 | Load shedding (MW) | 0 | 0 | 0 |
| | CHP generation (CHP1,CHP2) | 39.4 (0,39.4) | 39.4 (25,14.4) | 39.4 (39.4,0) |
| | Total cost ($) | 93411 | 95808 | 97890 |

TABLE VI
COMPARISON OF CASE 1 AND CASE 2 IN PERIODS 14 AND 24

| Period | Items | Pipelines | | | | | | |
|---|---|---|---|---|---|---|---|---|
| | | 1 | 2 | 3 | 4 | 5 | 6 | 7 |
| 14 | EF model | 49.97 | 45.32 | 26.96 | 4.59 | 18.33 | 0 | 18.27 |
| | Exact model | 50.07 | 45.36 | 26.98 | 4.60 | 18.34 | 0 | 18.28 |
| | Error | 0.22% | 0.09% | 0.06% | 0.35% | 0.06% | - | 0.04% |
| 24 | EF model | 0 | -3.59 | 20.86 | 3.56 | -24.51 | 38.67 | 14.14 |
| | Exact model | 0 | -3.60 | 20.88 | 3.57 | -24.53 | 38.79 | 14.15 |
| | Error | - | 0.38% | 0.07% | 0.43% | 0.11% | 0.30% | 0.04% |

### D. Accuracy of Energy Flow Model

In this paper, the energy flow model is utilized to consider the DHN reconfiguration, which approximates the heat loss. To validate its accuracy, the heat flow in each pipe with this model and the original exact DHS model for two typical hours (peak time of heat load at $t=24$ and valley time at $t=14$) are compared in Table VI. Because UC-CERHN problem with the exact DHS model is an intractable MINLP problem and constraint (10)-(13) cannot be formulated without known mass flow direction, the commitment variables are fixed in the exact model according to the result of the energy flow model. In Table VI, the results obtained by the two models are very close, and the maximum error is only 0.43%, which is acceptable in practice.

*E. Performance in Actual System in Jilin, China*

An actual system in Changchun, Jilin Province in China is used to test the scalability and effectiveness of the proposed UC-CERHN method. The system consists of a 319-bus EPS and five 8-node DHSs, three of which are operated by the same company and connected by tie pipes. There exist 5 CHP units, 3 heating boilers, 60 non-CHP units, and 34 wind farms in the system. The profiles of wind and load are real operating data, which are given in [22]. The results of Case 1 and Case 2 in the system are conducted for comparison. The results are shown in Table VII.

In Table VII, less CHP production is scheduled in Case 2 when DHN reconfiguration is enabled. The heat generation of CHP units is regulated with more flexibility in Case 2 by reallocating of heat power between different heat stations, which enhances the operational flexibility of the power system. Therefore, compared to those in Case 1, the total operation cost decreases by 1.3%, and the wind curtailment is reduced by 51.7% in Case 2, which proves the effectiveness of the proposed UC-CERHN method.

TABLE VII
OPERATION COST COMPARISON OF CASE 1 AND CASE 2 IN JILIN SYSTEM

| | $\sum_{t\in T}\sum_{i\in K^{TU}} C_{i,t}^{TU}$ | $\sum_{t\in T}\sum_{i\in K^{CHP}} C_{i,t}^{CHP}$ | $\sum_{t\in T}\sum_{i\in K^{WD}} C_{i,t}^{WD}$ | $\sum_{t\in T}\sum_{i\in K^{HB}} C_{i,t}^{HB}$ | Total cost |
|---|---|---|---|---|---|
| Case1 | 1863253 | 1333726 | 32304 | 28659 | 3257942 |
| Case2 | 1873416 | 1290321 | 15623 | 36514 | 3215874 |

## V. CONCLUSION

UC-CERHN is developed in this work to conduct a day-ahead commitment scheduling to coordinate the unit commitment of generators and valve operation in DHS. The reconfigurable DHS is characterized as a nonlinear and mixed-integer model. Also, an approximated generalized energy flow model is proposed to make the UC-CERHN problem tractable. The proposed UC-CERHN is capable of enhancing system energy efficiency and operational flexibility by utilizing the DHN reconfiguration effectively. Extensive case studies demonstrate that the DHN reconfiguration can provide a viable solution to improve the performance of power system in terms of wind power integration and congestion management. In addition, the accuracy of the proposed generalized energy flow model is verified, which can be also utilized in planning or regulation problems when the direction of mass flow varies.

A deterministic UC-CERHN model is applied in this paper. In our future work, the uncertainties of renewable energy and loads will be considered in our UC-CERHN model.